\documentclass[aps,prd,twocolumn,nofootinbib,dateno]{revtex4-2}
\usepackage{graphicx} 
\usepackage{comment}
\usepackage{hyperref}
\usepackage{natbib}
\usepackage{ragged2e} 
\usepackage{amsmath}
\usepackage{xcolor} 
\usepackage{soul} 
\soulregister\cite7

\usepackage{picture}
\usepackage{tikz}
\usetikzlibrary{calc}
\colorlet{lightgreen}{green!30}

\renewcommand{\hl}[1]{#1}
  
\begin{document}
\title{Strong CP Phase and Parity in the Hamiltonian Formalism}

\author{Ravi Kuchimanchi}
\email{raviparity@gmail.com }


\begin{abstract}

We show using the Hamiltonian formalism that if parity is a good symmetry of QCD, then the strong CP phase $\bar{\theta}$ must be $0$ or $\pi$. We find that for $P$ to be a physical symmetry, it must leave the Hilbert space $\mathcal{H}_\theta$ associated with the $\theta$-vacuum invariant ($P: \mathcal{H}_\theta \to \mathcal{H}_\theta$), which is possible only for $\theta = 0$ or $\pi$. We also show that  forming linear combinations of states from different $\theta$-sectors produces only classical statistical mixtures, consistent with superselection rules, confirming that $\mathcal{H}_\theta$ is the most general Hilbert space for the quantum theory. Furthermore, we demonstrate that requiring $[P,\Omega]=0$—where $\Omega$ is the generator of large gauge transformations—independently enforces $\bar{\theta}=0$ \hl{(mod $\pi$)}, and that for complex quark mass matrix $M$, if a generalized parity operator $\mathcal{P}$ is a symmetry, then the value of $\theta$ gets determined so that it exactly cancels $Arg Det M$, again giving $\bar{\theta}=0$ (mod $\pi$). These results establish the equivalence of the Hamiltonian and Lagrangian approaches to the strong CP problem.

\end{abstract}
\maketitle
\section{Introduction}
It is well known that in QCD, the non-perturbative tunneling between topological sectors $|n\rangle$ (where $n$ is the associated winding number) gives rise to the \(\theta\)-states~\cite{BELAVIN197585, PhysRevLett.37.172, CALLAN1976334}:
\begin{equation}
\left|\theta\right> = {{1}\over{\sqrt{2N}}}\sum_{n=-\infty}^{+\infty} e^{in\theta} \left|n\right>
\label{eq:thetavac}
\end{equation}
where $\sqrt{2N}$ is the normalization factor and the states $|n\rangle$ are obtained from the states with $|n=0\rangle$ by large gauge transformations 
\begin{equation}
\Omega |n\rangle = |{n+1}\rangle \implies \Omega |\theta \rangle = e^{-i\theta} |\theta \rangle.
\label{eq:omega}
\end{equation}

While we formulate the $\theta$-states following the familiar semi-classical picture in terms of instantons and winding-number sectors, the existence of $\theta$-dependence in QCD is a fully nonperturbative feature of Yang–Mills theory. As stressed in~\cite{RevModPhys.52.661}, the $\theta$ parameter arises in the exact quantum theory independent of any semi-classical approximation.

The phase \(\theta\) appears in the QCD Lagrangian via the term \(\theta G\tilde{G}\), and contributes to the strong CP phase 
\begin{equation}
    \bar{\theta} = \theta + Arg Det M
    \label{eq:barr}
\end{equation}
where M is the quark mass matrix,

 $\bar{\theta}$ contributes to the electric dipole moment of the neutron (nEDM) and is physically observable. However, despite decades of experimental effort, no evidence for nEDM has been found~\cite{PhysRevLett.124.081803}, and why the two terms on the right hand side of the above equation must cancel to yield $\bar{\theta} \approx 0$ to within a part in ten billion is the strong CP Problem, which can be resolved by introducing  axions~\cite{1977PhRvL..38.1440P, PhysRevLett.40.223,PhysRevLett.40.279} or through solutions based on discrete spacetime symmetries of the full Lorentz group, $P$ and/or $CP$ (or $T$)~\cite{Nelson:1983zb, PhysRevLett.53.329, PhysRevD.41.1286, Kuchimanchi:2010xs, bonnefoy2023colorful}. 

Recently, a proposal has appeared in the literature~\cite{kaplan2025solvestrongcpproblem} that implicitly assumes that the \(\theta\)-vacua can be coherently superposed, and the initial state of the universe is argued to be a linear combination of $|\theta \rangle$ states with different values of $\theta$.  The proposal draws analogies to Bloch wave superpositions in solid-state physics. 

If true, this would have consequences for the Standard Model (SM) itself, as it is argued in~\cite{kaplan2025solvestrongcpproblem}  that even the tuning for small $\bar{\theta}$ in eq.~(\ref{eq:barr}), as is done in SM,  would not work if the particular value of $\theta$ we see today is a random outcome of measurement from an initial superposition state of the universe. Moreover, it is claimed that solutions to the strong CP problem based on $P$ and $CP$ would also be nonviable. 

Independent of Ref~\cite{kaplan2025solvestrongcpproblem}, a key motivation for this work is that since \(G\tilde{G}\) is odd under both \(P\) and \(CP\), it is commonly argued that imposing these symmetries sets \(\theta = 0\) (or \(\pi\), since \(-\pi\) is identified with \(\pi\)) in the Lagrangian. However, how this arises in the Hamiltonian formalism has not been explored in the literature, and this can lead to the doubting or questioning of results obtained using the Lagrangian (as has happened in~\cite{kaplan2025solvestrongcpproblem}).

In this work, we first show that due to superselection rules, a linear combination across different $\theta$ sectors corresponds to a classical statistically mixed state rather than  a coherent quantum superposition, and therefore arguments made in \cite{kaplan2025solvestrongcpproblem} lack justification (see also~\cite{kuchimanchi2025paritysolvesstrongcp}).  

We then develop several new results within the Hamiltonian framework that determine the strong CP phase. In particular, we show that for $P$ to qualify as a physical symmetry, it must leave QCD's Hilbert space $\mathcal{H}_\theta$ invariant, which enforces $\bar{\theta}=0$ or $\pi$. We also point out that the commutation relation $[P,\Omega]=0$ with large gauge transformations provides an independent operator-level argument for $\bar{\theta}=0$. Finally, in the presence of complex fermion masses, we demonstrate that invariance under a generalized parity operator $\mathcal{P}$ fixes the value of $\theta$ so as to exactly cancel ${\rm Arg Det}M$, again ensuring $\bar{\theta}=0$ or $\pi$. Thus we establish the equivalence of the Hamiltonian and Lagrangian formalisms for the strong CP problem.


The paper is organized as follows:  In Section~\ref{sec:theta} we briefly revisit the relevant properties of the $\theta$-vacua and show that a superposition across different sectors corresponds to a mixed state rather than a pure quantum state.     In Section~\ref{sec:bloch} we argue that analogies with Bloch waves are misleading, since unlike in QCD, observations can connect different $k$ (analogous to $\theta$) states in condensed matter systems.  In Section~\ref{sec:thetabarr}, we show that $P$ (or $\mathcal{P}$ for complex fermion masses) sets $\bar{\theta} = 0$ or $\pi$ using the QCD Hamiltonian and the invariance of the Hilbert space $\mathcal{H}_\theta$, and alternatively obtain the same result using $[P,\Omega] = 0$ (or $[\mathcal{P}, \Omega]=0)$.    Section~\ref{sec:pcp} briefly discusses theories beyond the Standard Model that utilize $P$ and/or $CP$, and we conclude in Section~\ref{sec:conc}.



\section{\(\theta\)-states and Superselection in QCD}
\label{sec:theta}

 Gauge invariance of QCD requires that physical observables \(\mathcal{O}\) commute with the large gauge transformation operator \(\Omega\) of eq.~(\ref{eq:omega}), which implies:
\begin{equation}
    \langle n' | \mathcal{O} | n \rangle = \langle {n'+1} | \mathcal{O} | {n+1} \rangle = f({n'-n}),
\end{equation}

i.e., the matrix elements depend only on the difference ${m=n'-n}$. Evaluating the matrix element between \(\theta\)-states gives:
\begin{equation}
\langle \theta' | \mathcal{O} | \theta \rangle
= \left( \sum_{m} f(m) e^{-i m \theta'} \right)
\left( {{1}\over{2N}}\sum_{n} e^{i n (\theta - \theta')} \right) 
\label{eq:sumsum}
\end{equation}

Note that if $\theta = \theta'$ the second sum on the right hand side of eq.~(\ref{eq:sumsum})  counts the number of $|n\rangle$ states in $|\theta \rangle$ (see eq.~(\ref{eq:thetavac})) that are related to each other by large gauge transformations.  Since this sum is infinite we can regularize it by summing over $n$ from a large number $-N$ to $N$, which then cancels the normalization factor $2N$ in the denominator as $N \rightarrow \infty$, so that the term in the second brackets is $1$ if $\theta = \theta'$.\footnote{Note that we could have picked the normalization factor to be $\sqrt{2N+1}$ instead of $\sqrt{2N}$ in eq.~(\ref{eq:thetavac}), so that the term in the second round brackets of eq.~(\ref{eq:sumsum}) is $1$ for any $N$, if $\theta = \theta'$, and $n$ is summed from $-N$ to $N$. Both these normalization factors are the same since we take the limit $N\rightarrow \infty$.} While if $\theta \neq \theta'$, due to phase cancellations, this term  is zero as $N \rightarrow \infty$.

Moreover for observables that do not depend on instanton physics,  $f(m) =0$ if $m \neq 0$ (only instantons induce transitions between topologically different states), and therefore the first sum in eq.~(\ref{eq:sumsum}) is just $f(0)$.  Thus for all such observables $\langle \theta | \mathcal{O} | \theta \rangle = \langle n| \mathcal{O} |n \rangle$, independent of the value of $n$ and $\theta$.

The energy of states and neutron EDM depend on instanton physics and chiral anomaly, and therefore on $\theta$, and for these observables $f(m)$ is present even if $m \neq 0$. However even in this case the term in the second brackets vanishes if $\theta \neq \theta'$, as $N \rightarrow \infty$. 

Therefore, for all observables or gauge invariant operators  $\mathcal{O}$ we have~\cite{Shifman_2012, doi:10.1142/S0217751X02011357} 
\begin{equation}
   \langle \theta' | \mathcal{O} | \theta \rangle = 0 \ \ if \ \ \theta \neq \theta',
   \label{eq:zero}
\end{equation}

In~\cite{doi:10.1142/S0217751X02011357} the above is stated as $\langle \theta' | \hat{\mathcal{O}} | \theta \rangle = \delta_{\theta'\theta}$ with the interesting use of Kronecker delta function, and it is summarized as ``The different $\theta$ worlds are disconnected from each other." 

Thus, all gauge-invariant observables (and also the Hamiltonian),  are diagonal in \(\theta\), and different \(\theta\)-vacua are superselected.  Therefore we cannot transition from one theta vacuum to another by time evolution or by collapsing the wavefunction through observations.

The above superselection also implies that we cannot have a meaningful superposition of states that have
different values of $\theta$, as they are in different Hilbert
spaces  and there can be no interference or coherence between such
states~\cite{Gomez}. 

Note that even if one were to formally write a linear combination of different 
$\theta$ states, due to eq.~(\ref{eq:zero}) this would in fact correspond to a classical statistical mixture of distinct  $\theta$ sectors, and not a coherent quantum superposition ~\cite{PhysRev.88.101, Wightman1995,Giulini2009} (see for example discussion around equations $(1)$ and $(2)$ of~\cite{Giulini2009}). Consequently, a measurement of $\theta$ would not collapse the wavefunction across sectors, but merely reveals which sector the system was in to begin with. Thus, the reasoning in~\cite{kaplan2025solvestrongcpproblem}, that works only if $\theta$ is a random outcome of measurement of a pure quantum state, is not viable.

Likewise, because of  eq.~(\ref{eq:zero}) and Refs.~\cite{PhysRev.88.101, Wightman1995,Giulini2009}, a state such as $\left(|\theta\rangle + |-\theta \rangle \right)/\sqrt{2}$ is not a pure quantum state that is symmetric under an operator such as parity ($P$) that reverses the sign of $\theta$, but is simply a statistical mixture of two quantum states, neither of which is $P$-symmetric unless $\theta = 0$ or $\pi$. In fact, replacing the plus sign between the two states with a minus sign (or any other relative phase) leads to the same statistical mixture or  mixed state,  since the relative phases between different $\theta$-sectors have no physical meaning~\footnote{On the other hand, Ref.~\cite{kaplan2025solvestrongcpproblem} states ``...a viewpoint that one cannot form superpositions of states in different superselection sectors (for example, to argue that the only $P$ symmetric states are $\theta = 0$ or $\theta = \pi$) is itself an additional assumption, not a consequence of any fundamental quantum mechanical principle." However coherent  superpositions across superselection sectors are not possible as shown in Refs.\cite{PhysRev.88.101, Wightman1995,Giulini2009}, and $\theta =0$ or $\pi$  are the only $P$-symmetric states of the quantum theory.}. 

In essence, states with different values of $\theta$ are separated by superselection rules that do not allow their coherent quantum superposition.



The Hilbert space can be thought of as a direct sum over superselection sectors:
\begin{equation}
\mathcal{H} = \bigoplus_{{\theta'} \in (-\pi,\pi]} \mathcal{H}_{\theta'},
\end{equation}
where the Hilbert space of QCD is \(\mathcal{H}_\theta\) with a fixed value of the parameter \(\theta\). All the states generated from the theta vacuum with this value of $\theta$ lie within $\mathcal{H}_\theta$. 

Note that $\mathcal{H}_\theta$ is  the most general Hilbert space for the quantum theory to reside in, as there can only be a classical statistical mixture between states residing in $\mathcal{H}_\theta$ and $\mathcal{H}_{\theta'}$ (for $\theta' \neq \theta$).

We can now use the standard delta function normalization for the second sum in eq.~(\ref{eq:sumsum}) so that with $N = \pi$ we have
\begin{equation}
\langle \theta' | \mathcal{O} | \theta \rangle = \delta(\theta - \theta') \sum_m f(m) e^{i m \theta},
\end{equation}
which vanishes unless \(\theta = \theta'\), and leads to the same eq.~(\ref{eq:zero}) and conclusions stemming from it. 

While this normalization is common in QFT texts and lecture notes, we first derived a simpler, more transparent normalization by summing over the number of $|n\rangle$ states in $|\theta \rangle$ and taking the limit  $N \rightarrow \infty$. This approach has the advantage of working entirely within QCD's Hilbert space $\mathcal{H}_\theta$, without needing to invoke the direct sum over $\theta$-sectors or delta-function normalization.


\section{Comparison with Bloch Bands}
\label{sec:bloch}

Reference~\cite{kaplan2025solvestrongcpproblem} draws analogy with a periodic potential, where the crystal momentum \(k\) plays a role similar to \(\theta\), to argue that QCD has a Bloch-like band structure, consisting of states with different values of $\theta$. 

However, in quantum mechanics with periodic potential,  operators like position \(x\) do not commute with the translation symmetry operator $T$   (which translates the system by one period of the potential and is analogous to $\Omega$ that shifts by one winding number), and measurements in real space can therefore project onto superpositions of \(k\)-states. In QCD no gauge-invariant observable can connect or interfere between different \(\theta\)-states. Thus, the analogy breaks down when we consider physical observables.

That is $\langle k'|x|k\rangle \neq 0$ for $k \neq k'$, for observables like the position operator $x$, and therefore there is no analog  of QCD's eq.~(\ref{eq:zero}) that applies to all observables, in the problem of periodic potential and Bloch waves. And there is no analogous Bloch-like band structure in QCD.

As emphasized by Jackiw in~\cite{RevModPhys.52.661}, the quantum mechanical analogy of the Bloch waves (where $\theta$ is like $k$) has an important limitation: 
\begin{quote}
``While the quantum-mechanical analogy is a good one,
there is an important difference from the Yang-Mills
theory. In the crystal, all values of $\theta$ are attainable,
and $\theta$ is a measure of the energy in an energy band.
In the Yang-Mills theory, when gauge-invariant quantities are considered, $\theta$ cannot change, so that even
though one can imagine states with different $\theta$, a complete physical theory is characterized by a single,
unique, but as yet undetermined angle. (It is as if in
the crystal example, all physical observables were
translationally invariant; in that case only one state
per band would be physically realizable.)"
\end{quote}

Tong makes a similar observation in Section 2.2.3 of his lectures~\cite{Tong:2000}, drawing the contrast between the band of states in the Bloch case and a single $\theta$ state in QCD:
\begin{quote}
    ``For the particle in a potential, all the states \(|k\rangle\) lie in the Hilbert space. Indeed, the spectrum famously forms a band labeled by \(k\). In contrast, in Yang-Mills theory there is only a single state: each theory has a specific \(\theta\) which picks out one state from the band. This can be traced to the different interpretation of the group generators. The translation operator for a particle is a genuine symmetry, moving one physical state to another. In contrast, the topologically non-trivial gauge transformation \(\Omega\) is, like all gauge transformations, a redundancy: it relates physically identical states, albeit up to a phase.''\\
    \end{quote}

    Since the theory has a specific value of $\theta$, we now turn to the question of how this value — and the corresponding Hilbert space $\mathcal{H_\theta}$  — is determined when parity is a good symmetry. A similar analysis can be carried out using $CP$ symmetry instead.

\section{Parity and the Selection of $\bar{\theta} = 0$ or $\pi$ from the Hamiltonian Perspective}
\label{sec:thetabarr}
Generally $\theta$ is treated as a parameter in the Lagrangian that vanishes due to $P$ or $CP$ symmetries. However it hasn't been shown so far using operators, commutation relations and QCD's Hilbert space that they act on, that these symmetries determine that $\bar{\theta} = 0$ or $\pi$. We now proceed to show this, which also demonstrates the equivalence of the Hamiltonian and the Lagrangian formalisms for the strong CP problem.

The QCD Hamiltonian \(H\) for a single quark flavor contains the mass term:
\begin{equation}
 H_{\text{mass}} =   m(\bar{\psi}_L \psi_R) + m^\star(\bar{\psi}_R \psi_L) 
 \label{eq:m}
\end{equation}

This term respects parity ($P$) if  $PHP^{-1} = H$, or equivalently,  $[P,H]=0$. Since under  \(P\) we have, $x \rightarrow -x$ and $\psi_L \leftrightarrow \psi_R$,  invariance of the mass term requires, $m = m^\star$, i.e. that $m$ is real.

Since there are superselection rules,  for \(P\) to be a physical symmetry, it must leave the  Hilbert space \(\mathcal{H}_\theta\) on which it acts invariant, i.e., \(P: \mathcal{H}_\theta \rightarrow \mathcal{H}_\theta\). Parity inverts winding numbers, $P|n\rangle = |{-n\rangle}$, and from eq.~(\ref{eq:thetavac}) we can see that \(P\) maps \(|\theta\rangle \to |{-\theta}\rangle\) (that is, $\mathcal{H}_\theta \rightarrow \mathcal{H}_{-\theta}$).  Therefore, \(\mathcal{H}_\theta\)  is invariant under $P$  only if \(\theta = -\theta \mod 2\pi\), i.e., \(\theta = 0\) or \(\pi\). 

 Note that if this were not the case, then $P$ would be a map from one theory to a different theory (with a different value of $\theta$), rather than a good symmetry within the same theory.  

A further understanding arises when we ask, since P is a good symmetry shouldn't it commute with gauge transformations? 

Since parity flips the winding number of gauge configurations,  the large gauge transformation operator \(\Omega\) satisfies \(P \Omega P^{-1} = \Omega^{-1}\), implying \([P, \Omega] \ne 0\) in general. However, $P\Omega|\theta\rangle = \Omega P |\theta \rangle$ if  \(e^{-i\theta} = e^{i\theta}\).  That is,  $[P, \Omega] = 0$ on \(\mathcal{H}_\theta\) if $\theta = 0$ or $\pi$.

Since $P$ appears to lead to two conditions, namely,  $\theta=0$ and $m$ being real (equivalently, $Arg Det M = 0$), more clarity may be needed on what $P$ is really implying. Therefore we now do a more general analysis that shows that what is determined is that the sum in eq.~(\ref{eq:scp}), $\bar{\theta} = 0~or~\pi$.   

Now consider the more general case of eq.~(\ref{eq:m}) where \(m = |m| e^{i\alpha}\) and is complex.  We define a generalized parity transformation $\mathcal{P} = P \cdot U_A(-\alpha)$, where \(U_A({-\alpha})\) is an axial rotation that transforms $\psi_L \rightarrow e^{i\alpha}\psi_L$ and $\psi_R \rightarrow e^{-i\alpha}\psi_R$. It is easy to check that the mass term in eq.~(\ref{eq:m})  is invariant under $\mathcal{P}$. 

Moreover \(\mathcal{P}\) maps, 
\begin{equation}
\mathcal{H}_\theta \rightarrow \mathcal{H}_{-\theta - 2\alpha},
\label{eq:map}
\end{equation}
where the $\alpha$ dependence is due to the chiral anomaly,  
\begin{equation}
U_A{(-\alpha)}|\theta\rangle = |\theta + 2 \alpha\rangle.
\label{eq:axial}
\end{equation}
Requiring as before that $\mathcal{P}$ leave the Hilbert space  $\mathcal{H}_\theta$ invariant implies $\theta = -\theta-2\alpha$ mod $2 \pi$ in eq.~(\ref{eq:map}), or $\theta +\alpha = 0$ mod $\pi$.  From eq.~(\ref{eq:barr}) we see that this sum is the Strong CP phase and we obtain, 
\begin{equation}
\bar{\theta} =\theta + \alpha =  0 \mod \pi
\label{eq:scp}
\end{equation}
where we have used $Arg Det M = \alpha$.

We stress that this condition does not require all states in the theory to be parity eigenstates. For example, one could consider a state consisting only of left-handed (or its parity counterpart only right handed) quarks. Rather, the requirement is that the Hilbert space of the theory  be closed under parity (or generalized parity), meaning that acting with $P$ (or $\mathcal{P}$) maps any physical state to another state within the same space --- this means, within the same \(\mathcal{H}_\theta\) for the QCD gauge fields. 

Alternatively, we can find the Hilbert space $\mathcal{H}_\theta$ in which \(\mathcal{P}\) commutes with \(\Omega\). 
\begin{equation}
    \left[\mathcal{P},  \Omega\right] |\theta\rangle = \left(e^{-i\theta}  - e^{i(\theta + 2\alpha)}\right) |{-\theta-2\alpha\rangle}
\end{equation}
where we have used eq.~(\ref{eq:axial}).

The right hand side of the above equation vanishes if $-\theta = \theta + 2\alpha$ mod $2\pi$ and we once again obtain 
\begin{equation}
    \theta = -\alpha ~mod~ \pi~~  if~~ [\mathcal{P}, \Omega] = 0~~on~~\mathcal{H}_\theta
    \end{equation}
    and therefore also eq.~(\ref{eq:scp}).

    Thus, \(\mathcal{P}\) is a valid symmetry that commutes with \(\Omega\) only when \(\bar{\theta} = 0\) or \(\pi\).

\section{Relevance beyond Standard Model}
\label{sec:pcp}
In theories such as the left-right symmetric model~\cite{PhysRevD.10.275,*PhysRevD.11.566,*Senjanovic:1975rk,PhysRevD.44.837, Duka:1999uc} based on $SU(3)_c \times SU(2)_L \times SU(2)_R \times U(1)_{B-L} \times P$, the direct product indicates that $P$ commutes with all the gauge symmetries, including the large gauge transformation $\Omega$ of QCD.  And similar is the case with models based on $SU(3)_C \times SU(2)_L \times U(1)_Y \times CP$, where $CP$ commutes.

Our analysis of the previous section then tells us that the standard practice of defining $P$ (or $CP$) such that the Hamiltonian (or the Higgs potential) is invariant (for example Hermitian Yukawa matrices for the left right symmetric model), will determine $\theta$ so that $\bar{\theta} =0$ mod $\pi$,  if P (or CP) are not broken spontaneously. This is consistent with considering $\theta$ as a parameter of the Lagrangian and setting it to zero (mod $\pi$) due to P.  A non-zero value of $\bar{\theta}$ is generated from Higgs VEVs only on spontaneous breaking of these discrete space-time symmetries. 


Note also that our conclusions apply not only to models that solve the strong CP problem using $P$ and/or $CP$, but also to several models that impose these symmetries without solving the strong CP problem.  Even in the latter case, the $\bar{\theta}$ is only generated due to the spontaneous breaking of these discrete symmetries and so is calculable from   the parameters of the theory.  Imposing the experimental constraint that $\bar{\theta} \leq 10^{-10}$, would then restrict the parameter space of these theories, and can lead to predictions for future experiments.  An example is the minimal left-right symmetric model with parity (LRSM) that doesn't by itself solve the strong CP problem, but there are consequences for it which can be tested~\cite{Kuchimanchi_2015}.

Experimental evidence for the LRSM model based on  $SU(3)_c \times SU(2)_L \times SU(2)_R \times U(1)_{B-L} \times P$, may come in the next decade or two  from the neutrino experiments DUNE and Hyper-K. It is now  known that  CP violation in the leptonic sector can radiatively generate a large strong CP phase in this model, due to one loop RGE running of a dimensionless quartic coupling parameter of the Higgs potential~\cite{Kuchimanchi_2015}. Since the key parameter is dimensionless, the dependence on large mass scales is logarithmic $\sim ln(v_R/M_{Pl})$ and there is no suppression of the strong CP phase by factors like $(v_{wk}/{v_R})^2$, where $v_R$ is the $SU(2)_R \times U(1)_{B-L}$ breaking scale  and $M_{pl}, v_{wk}$ are Planck and weak scales.   Therefore leptonic CP violation must be absent, and the CP phases in the PMNS matrix (including the Dirac phase $\delta_{CP}$ being probed by neutrino experiments) must be 0 or $\pi$ to well within a degree, in most of the parameter space. Thus LRSM anticipates that the neutrino experiments will find results consistent with $sin (\delta_{CP}) = 0$ in most of its parameter space, regardless of the high scale at which parity or $SU(2)_R \times U(1)_{B-L}$ are broken~\cite{Kuchimanchi_2015}.


We can solve the strong CP problem in the above LRSM by also including CP and adding a family of heavy quarks whose mixing with the usual quarks  generate the CKM phase when P and CP break spontaneously~\cite{Kuchimanchi:2010xs}.  In this minimal scenario no leptonic CP violation is generated~\cite{Kuchimanchi:2012te} at the tree-level, and once again we will expect null results from DUNE and Hyper-K for leptonic $\delta_{CP}$.  

Motivated by this experimental prediction for the leptonic sector, it was also realized in~\cite{Kuchimanchi:2012te} (see also~~\cite{PhysRevD.108.095023}) that even in Nelson-Barr models where CP and a shaping symmetry are used to solve the strong CP problem, these symmetries can be imposed such there is no leptonic CP violation generated at the tree-level after their spontaneous breaking (though they can also be imposed to generate leptonic CPV).   

Thus discovery in neutrino experiments of Dirac CP phase in PMNS matrix  consistent with $sin(\delta_{CP}) =0$ or $\pi$ can point towards $P$ and/or $CP$ symmetries being restored in the laws of nature.  Moreover, an absence of leptonic CP violation  cannot be explained by the axionic solution to the Strong CP problem.

The latest fit to global data from neutrino experiments
by Nu-fit 6.0 (2024)~\cite{esteban2024nufit60updatedglobalanalysis} has $\delta_{CP} = \pi$ to within one sigma of
its error bars for normal ordering of neutrino masses. We eagerly look forward to more data from T2K and NOvA, and to Hyper-K and DUNE experiments.

\section{Conclusions}
\label{sec:conc}

In this work we showed, using the Hamiltonian formalism, how parity (and analogously CP) symmetry determines the strong CP phase in QCD. 
Our analysis establishes several new results:

\begin{itemize}
    \item Requiring that for parity $P$ to be a good symmetry, it must leave the Hilbert space $\mathcal{H}_\theta$ invariant implies $\theta = 0$ or $\pi$ and $\text{Arg}\,\det M = 0 \ (\text{mod}\ \pi)$. 

    \item We further show that $[P,\Omega]=0$, where $\Omega$ is the generator of large gauge transformations, independently enforces $\bar{\theta}=0 \ (\text{mod}\ \pi)$. 

    \item For the case of complex fermion mass terms, we demonstrate that invariance under a generalized parity operator $\mathcal{P}$ (an axial rotation followed by $P$) fixes $\theta$ so that it precisely cancels $\text{Arg}\,\det M$, again giving $\bar{\theta}=0$ (mod $\pi$). 
\end{itemize}

In addition, we showed that formal linear combinations of states from different $\theta$-sectors correspond only to classical statistical mixtures, not to coherent quantum superpositions. This reinforces that $\mathcal{H}_\theta$ is the most general Hilbert space for QCD, in agreement with superselection rules. 

Together, these results show the equivalence of the Hamiltonian and Lagrangian approaches to the strong CP problem. They not only show P and CP-based solutions are viable, but also contribute significant new understanding of the role of symmetry, large gauge transformations, and generalized parity in constraining $\bar{\theta}$.





\vspace{0.25in}

\section{Note added}
Upon completion of this work, we became aware of the recent preprint~\cite{Vecchi:2025qie}, which includes a comment section consistent with the conclusions presented here and in~\cite{kuchimanchi2025paritysolvesstrongcp}, and in variance with those of~\cite{kaplan2025solvestrongcpproblem}.

\section{Acknowledgments}
I thank Rabi Mohapatra and Kaustubh Agashe for insightful discussions.

\bibliography{main}


\end{document}